\documentstyle[12pt]{procsla}
\begin{document}

\begin{center}
{\bf EXCLUSIVE NONLEPTONIC BOTTOM TO CHARM BARYON \\
DECAYS INCLUDING NONFACTORIZABLE CONTRIBUTIONS}\\

\vspace*{1cm}
M. A. IVANOV$^{1}$, J. G. K\"{O}RNER$^2$, V. E. LYUBOVITSKIJ$^{1,3}$ and
A. G. RUSETSKY$^{1,4}$
\\

\vspace*{.3cm}
{\em $^1$ Bogoliubov Laboratory of Theoretical Physics, JINR},

{\em Dubna (Moscow region), 141980 Russia}\\

{\em $^2$ Johannes Gutenberg-Universit\"{a}t,
Institut f\"{u}r Physik, D-55099 Mainz, Germany}

{\em $^3$ Department of Physics, Tomsk State University,
634050 Tomsk, Russia}

{\em $^4$ IHEP, Tbilisi State University, 380086 Tbilisi, Georgia}

\end{center}

\vspace*{1cm}
\baselineskip 15pt
\begin{center}
\parbox{15cm}{\small
Exclusive nonleptonic decays of bottom baryons to charm baryons and
pseudoscalar light mesons are analyzed within
a relativistic three-quark model. We include factorizing as well
as nonfactorizing contributions to the decay amplitudes. The total
contribution of the nonfactorizing diagrams amount up to $30~\%$
of the factorizing contributions in amplitude. We present detailed
predictions for rates and asymmetry parameters.
\normalsize
}
\end{center}

\vspace*{.5cm}

\section{Introduction}
\baselineskip 20pt
Recently  there has been much progress in the experimental
analysis of decays of heavy baryons \cite{PDG}$^-$\cite{CDF}.
From a number of experimental collaborations (ALEPH, ARGUS,
ACCMOR, CLEO, OPAL, etc.) there exist many results on the
mass spectrum, lifetimes, branching ratios and asymmetry
parameters of heavy baryon decays. In the near future one
can expect large quantities of new data on
exclusive nonleptonic bottom baryon  decays which calls for
a comprehensive theoretical analysis of these decays.

The analysis of nonleptonic heavy baryon decays is complicated
by the necessity of having to include nonfactorizing contributions.
One thus has to go beyond the factorization approximation which
has proved quite useful in the analysis of the exclusive nonleptonic
decays of heavy mesons~\cite{Bauer}. There have been some theoretical
attempts to analyze nonleptonic heavy baryon decays using the
factorizing contributions alone~\cite{Cheng3}, the argument being that
$W$-exchange contributions can be neglected in analogy to the power
suppressed $W$-exchange contributions in the inclusive nonleptonic
decays of heavy baryons. One might even be tempted to drop the
nonfactorizing contributions on account of the fact that they are
superficially proportional to $1/N_c$. However, since $N_c$-baryons
contain $N_c$ quarks an extra combinatorial factor proportional to
$N_c$ appears in the amplitudes  which cancels the explicit diagrammatic
$1/N_c$ factor~\cite{Kramer,SantaFe}. There is now ample empirical
evidences in the $c\rightarrow s$ sector that the nonfactorizing
diagrams cannot be neglected. For example, the two observed decays
$\Lambda_c^+\to \Xi^0 K^+$ and $\Lambda_c^+ \to \Sigma \pi$  can only
proceed via the nonfactorizing diagrams. Their sizeable observed branching
ratios may thus serve to obtain a measure of the size of the nonfactorizing
contributions.

In this paper we present a complete analysis of the exclusive
nonleptonic decays of bottom baryons $(\frac{1}{2}^+)$ into charm
baryons $(\frac{1}{2}^+)$ and light pseudoscalar mesons ($0^-$)
within the so-called {\it Lagrangian spectator model}
which has been developed from relativistic quark
model~\cite{Anikin}$^-$\cite{Mainz} and  may be viewed as the
generalization of the spectator quark model~\cite{Hussain}.
Both factorizing and nonfactorizing diagrams
can be evaluated in a self-consistent manner within this approach.
We calculate branching ratios and the asymmetry parameters of all
the decays in this class. The main result of our analysis is that
the nonfactorizing contributions are important also in the
$b\rightarrow c$ sector and cannot be neglected. In the decays with
both factorizing and nonfactorizing contributions the nonfactorizing
diagrams contribute destructively and can amount up to 30~\% of the
nonfactorizing contribution in amplitude. Some of the decay channels
as e.g. $\Lambda_b^0\to \Sigma_c^+ \pi^-$ and
$\Xi_b^0\to\Sigma_c^+ K^-$ have no factorizing contribution but are
predicted to occur in our approach albeit with a small branching fraction.
It would be important to have an experimental confirmation of
this prediction.

\section{Lagrangian spectator model}
\baselineskip 20pt

In this paper we will use a relativistic quark model developed
in~\cite{Anikin,PIGG} to calculate nonleptonic decays of heavy baryons.
This model has been successfully applied to the description of the
electromagnetic properties of nucleons~\cite{PSI} and has been extended
to an analysis of the semileptonic decays of heavy baryons~\cite{Mainz}.
Since the evaluation of the nonleptonic decay amplitudes involves
three-loop diagrams with nonlocal vertices, we shall make some
simplifying assumptions which, on the one hand, have a clear physics
motivation, and, on the other hand, allow one to evaluate both the
factorizing and the nonfactorizing contributions to nonleptonic baryon
decays.

Let us begin by recalling some of the crucial points of the approach
developed in~\cite{Anikin}$^-$\cite{Mainz}. We consider the hadron to be a
bound state of relativistic constituent quarks. The coupling of the
hadrons to their constituent quarks is described by an interaction
Lagrangian with an effective vertex function characterizing the momentum
distribution of the constituents.

The Lagrangian describing the interaction of baryons with the
three-quark current is written as
\begin{eqnarray}\label{strong}
{\cal L}_B^{\rm int}(x)&=&g_B\bar B(x)
\hspace*{-.1cm}\int \hspace*{-.1cm}dy_1\hspace*{-.1cm}\int
\hspace*{-.1cm}dy_2\hspace*{-.1cm}\int \hspace*{-.1cm}dy_3 \,
\delta\left(x-\frac{\sum\limits_i m_iy_i}{\sum\limits_i m_i}\right)
F\left(\frac{\Lambda_B^2}{18}\sum\limits_{i<j}(y_i-y_j)^2\right)
\nonumber \\
&\times&J_B(y_1,y_2,y_3)+{\rm h.c.}
\end{eqnarray}
where
$J_B(y_1,y_2,y_3)=\Gamma_1 q^{a_1}(y_1)q^{a_2}(y_2)C\Gamma_2 q^{a_3}(y_3)
\varepsilon^{a_1a_2a_3}$ is the three-quark current with the quantum
numbers of the baryon $B$. The spatial four-coordinates $y_i$ (i=1,2,3)
of the quarks are expressed through the c.m. coordinate $(x)$ and the
relative Jacobi coordinates $(\xi_1,\xi_2)$ \cite{Mainz}. The
$\Gamma_{1,2}$ are strings of Dirac matrices, $C$ is the charge
conjugation matrix, and $a_i$ are colour indices.

The spin-flavour structure of heavy-light baryon currents has been
studied in detail in the papers~\cite{Mainz,Shuryak,Grozin}. It was
shown that in the heavy quark limit there are two currents for
$\Lambda$-type baryons containing a light diquark with spin zero
and two currents for $\Omega$-type baryons having a light
diquark with spin 1. In ref.\cite{Mainz} we have shown that in the
heavy quark limit the heavy quark factorizes from the light
degrees of freedom. Then the Lagrangian which describes the
interaction of $\Lambda_Q$-baryon with the constituent quarks may be
written  as
\begin{eqnarray}\label{lagr_LB}
{\cal L}_{\Lambda_Q}^{\rm int}(x)&=&g_{\Lambda_Q}\bar \Lambda_Q(x)
\Gamma_1 Q^a(x)\int d\xi_1 \int d\xi_2
F(\Lambda_{B_Q}^2\cdot[\xi_1^2+\xi_2^2])\\
&\times&u^b(x+3\xi_1-\xi_2\sqrt{3})C\Gamma_2d^c(x+3\xi_1+\xi_2\sqrt{3})
\varepsilon^{abc}+{\rm h.c.}
\nonumber
\end{eqnarray}
where $\Gamma_1\otimes C\Gamma_2=I\otimes C\gamma^5$ or
$\gamma_\mu \otimes C\gamma^\mu\gamma^5$. The explicit form of
the interaction Lagrangians of light baryons with their constituents
can be found in Ref.~\cite{Mainz}.

We are modelling the effective vertex function $F$ in Eq.~(\ref{strong})
by the Gaussian shape factor $F(k^2_E)=\exp(-k^2_E/\Lambda_B^2$) which
falls off sufficiently fast in the Euclidean region to provide for the
ultraviolet convergence of the matrix elements. It was shown in~\cite{Mainz}
that the requirement of the correct unit normalization of the baryonic
IW-functions $\zeta(\omega)$ and $\xi_1(\omega)$ at zero recoil
$\omega=1$ imposes the condition $\Lambda_{B_b}=\Lambda_{B_c}$. For light
baryons we introduce the cutoff parameter $\Lambda_{B_q}$.
The cutoff parameters $\Lambda_{B_Q} = \Lambda_{B_b} = \Lambda_{B_c}$ and
$\Lambda_{B_q}$ are the adjustable parameters.

The Lagrangian spectator model has been derived from
the relativistic quark model~\cite{Anikin}$^-$\cite{Mainz}. It aims to
reproduce the spin amplitude structure of the spectator (or static quark)
model analysis~\cite{Kramer}. This can be achieved by assigning the
projector $V_+=(1\,\, + \not\! \hspace*{-.5mm}v)/2$ to each light quark field
in the baryon-quark vertex, and by using the static approximation for
$u$, $d$ and $s$ quark propagators
\begin{eqnarray}\label{static_prop}
<0|T\{q(x)\bar q(y)\}|0> = \frac{1}{\Lambda_q} \,\, \delta^{(4)}(x-y)
\end{eqnarray}
\noindent where $\Lambda_q$ is the free parameter having the dimension
of mass. We choose this parameter to have the same value $\Lambda$ for
$u$ and $d$ quarks and a different value $\Lambda_s$ for the strange quark.
Note that in this model the above
two options of pseudoscalar and axial currents for the $\Lambda$-type
baryon and two options of vector and tensor currents for the
$\Omega$-type baryon become equivalent in the Lagrangian spectator
model. It may be seen by using the equation of motion
$\bar B_Q(x) \not\! \hspace*{-.1cm}v = \bar B_Q(x)$. The baryon-quark
coupling constants in Eqs.(\ref{strong},\ref{lagr_LB}) are fixed from
{\it the compositeness condition} which is equivalent to the unit
normalization of the elastic baryon form factor at the origin~\cite{Mainz}.

In this paper we also assume for simplicity that the mesons are point-like
objects, i.e. their interaction with the constituent quarks are described
by a local nonderivative Lagrangian.
In other words, we choose the effective meson vertex functions to be
constant in momentum space. This is a reliable approximation for the
light mesons. For heavy mesons we expect that form factor effects in the
meson vertex become important. This prevents us from extending the present
approach to cases with heavy mesons in the final states, such as
$\Lambda_b^0\rightarrow\Lambda_c^++D_s^-$. In general the form factor
effects in the decays involving heavy mesons in the final state
are expected to suppress their rates relative to those obtained from
a point-like vertex. Exclusive nonleptonic bottom baryon decays involving
heavy mesons form the subject of a separate piece of work.

For the heavy quark propagator $S_Q$ we will use the leading term in the
inverse mass expansion. Suppose $p=M_{B_Q}v$ is the heavy baryon momentum.
We introduce a set the binding energy parameters
$\bar\Lambda_{\{q_1q_2\}}=M_{\{Qq_1q_2\}}-m_Q$
which parametrize the difference between the heavy baryon mass
$M_{\{Qq_1q_2\}}\equiv M_{B_Q}$
and the heavy quark mass. Keeping in mind that the vertex function
falls off sufficiently fast such that the condition
$|k|<<m_Q$ holds ($k$ is the virtual momentum of light quarks) one has
\begin{eqnarray}\label{Sheavy}
S_Q(p+k)&=&{1\over m_Q - (\not\! p \,\, + \not\! k)}=
S_v(k,\bar\Lambda_{\{q_1q_2\}})+O\biggl(\frac{1}{m_Q}\biggr)
\nonumber\\[5mm]
& &S_v(k,\bar\Lambda_{\{q_1q_2\}}) = - \frac{(1+\not\! v)}
{2(v\cdot k + \bar\Lambda_{\{q_1q_2\}})}
\end{eqnarray}
In what follows we will assume that
$\bar\Lambda\equiv\bar\Lambda_{\{uu\}}=\bar\Lambda_{\{dd\}}=
\bar\Lambda_{\{du\}}$,
$\bar\Lambda_{s}\equiv\bar\Lambda_{\{us\}}=\bar\Lambda_{\{ds\}}$.
Thus there are altogether three independent binding energy parameters
$\bar\Lambda$, $\bar\Lambda_{s}$, and $\bar\Lambda_{\{ss\}}$
in our approach.

In the Lagrangian Spectator Model the leptonic coupling
constants $f_\pi$ and $f_K$ are determined by the integrals
\begin{eqnarray}\label{fpifk}
f_\pi = \frac{N_cg_\pi}{4\pi^2} \frac{1}{M_\pi \Lambda^2}
\int_{reg} \frac{d^4 k}{\pi^2}, \hspace*{2cm}
f_K   = \frac{N_cg_K}{4\pi^2} \frac{1}{M_K \Lambda\Lambda_s}
\int_{reg} \frac{d^4 k}{\pi^2}\label{fpik}
\end{eqnarray}
\noindent The meson coupling constants $g_\pi$ and $g_K$ in
Eq.(\ref{fpifk}) are determined from
{\it the compositeness condition}~\cite{Mainz} which reads
\begin{eqnarray}
1 = \frac{N_cg_\pi^2}{4\pi^2} \frac{1}{M_\pi^2 \Lambda^2}
\int_{reg} \frac{d^4 k}{\pi^2}, \hspace*{2cm}
1 = \frac{N_cg_K^2}{4\pi^2} \frac{1}{M_K^2 \Lambda\Lambda_s}
\int_{reg} \frac{d^4 k}{\pi^2}\label{gpik}
\end{eqnarray}
\noindent Equations (\ref{fpik}) and (\ref{gpik})
contain the ultraviolet divergence since the mesons in our scheme
are point-like objects. To regularize these quantities we introduce
an ultraviolet cutoff parameter
$\Lambda_{cut}=\Lambda_{q_1}\Lambda_{q_2}/(\Lambda_{q_1}+\Lambda_{q_2})$
with the same $\Lambda_{q_i}$ as in Eq.(\ref{static_prop}). Here $q_i$
corresponds to the flavour of the light quark being the constituent.
After that we get
\begin{eqnarray}\label{lepcon}
f_\pi = \frac{\sqrt{N_c}}{8\pi} \Lambda, \hspace*{2cm}
f_K = \frac{\sqrt{N_c}}{2\pi} \frac{(\Lambda\Lambda_s)^{3/2}}
{(\Lambda+\Lambda_s)^2}
\end{eqnarray}
\noindent Substituting experimental values for $f_\pi$ = 131 MeV
and $f_K$ = 160 MeV in Eqs.(\ref{lepcon}) we obtain $\Lambda$=1.90 GeV
and $\Lambda_s$=3.29 GeV.

Thus, there is the following set of adjustable parameters in our model:
the cutoff parameters  $\Lambda_{B_q}$ and $\Lambda_{B_Q}$,
and the set of binding energy parameters $\bar\Lambda$,
$\bar\Lambda_s$ and $\bar\Lambda_{\{ss\}}$.

\section{Nonleptonic transition matrix elements}

The weak nonleptonic decays of bottom and charm baryons are described
by the diagrams Fig.1. Diagram I corresponds to the so-called factorizing
contribution while  the diagrams II and III correspond to the nonfactorizing
contributions.
\footnote
{In the terminology of~\cite{Cheng3} diagram I corresponds to factorizable
external and internal W-emission, IIa to nonfactorizable internal W-emission
and IIb and III to nonfactorizable W-exchange.}
The vertices $O_\mu\bullet \bullet O_\mu$ arise from the
standard effective four-fermion Lagrangian \cite{Buras} which for
$b\to c \bar u d$ transitions has the form
\begin{eqnarray}\label{eff}
\hspace*{-0.8cm}{\cal L}_{\rm eff}=\frac{G_F}{\sqrt{2}}V_{cb}V^\dagger_{ud}
[c_1(\bar c^{a_1} O_\mu b^{a_1}) (\bar d^{a_2} O_\mu u^{a_2})
+ c_2(\bar c^{a_1} O_\mu b^{a_2}) (\bar d^{a_2} O_\mu u^{a_1})] + {\rm h.c.}
\end{eqnarray}
Here $c_1$ and $c_2$ are the short distance Wilson coefficients
\cite{Buras}.
It is well-known that the factorizing contributions are proportional to the
two linear combinations $a_1=c_1 + c_2/N_c$ and $a_2=c_2 + c_1/N_c$
where $N_c$ is the number of colours.
An analysis of the
nonleptonic decays of $B$ mesons gives the values:
$a_1 \approx 1.05 \pm 0.10$ and $a_2 \approx 0.25 \pm 0.05$ \cite{Buras}.

After some straightforward calculations the matrix element describing
exclusive  weak nonleptonic decays of bottom baryons can be written
as~\cite{Kramer}
\footnote
{We employ the notation
\[ \gamma_5=\left(
\begin{array}{cc}
0  & -I \\
-I & 0
\end{array} \right)\]
}

\begin{equation}\label{Minv}
M= M_{\rm I}+M_{\rm IIa}+M_{\rm IIb}+M_{\rm III}\equiv A-\gamma_5 B
\end{equation}
where the amplitudes $M_{\rm I}$, $M_{\rm IIa}$, $M_{\rm IIb}$,
and $M_{\rm III}$ correspond to the contributions of
diagrams I, IIa, IIb, and III in Fig.1, respectively. One has

\noindent
\underline{Factorizing contribution:}
\begin{equation}\label{fac1}
{\rm Diagram \;\; I:} \hspace{.5cm}
M_{\rm I}=c_W\chi_{\pm}f_P\frac{Q_+}{4M_1M_2}
\biggl(M_-\ell^-_{FD}-M_+\ell^+_{FD} \cdot \gamma^5\biggr) f(\omega)
\end{equation}

\vspace*{1cm}
\noindent
\underline{Nonfactorizing contributions}
\begin{eqnarray}
& &{\rm Diagram \;\; IIa:} \hspace{.5cm} M_{\rm IIa}=
c_Wc_-\frac{H_2(\omega)}{4M_1M_2}
\biggl(P_+\ell^{P^+}_{II_a}-Q_+\ell^{Q^+}_{II_a} \cdot \gamma^5\biggr) M_1
\label{fac2a} \\
& &{\rm Diagram \;\; IIb:} \hspace{.5cm}
M_{\rm IIb}=c_Wc_-\frac{H_2(\omega)}{4M_1M_2}
\biggl(D_+\ell^{D^+}_{II_b}-Q_+\ell^{Q^+}_{II_b} \cdot \gamma^5 \biggr) M_2
\label{fac2b} \\
& &{\rm Diagram \;\; III:} \hspace{.5cm}
M_{\rm III}=c_Wc_-\frac{H_3(\omega)}{4M_1M_2}
\sum_{i=1}^3M_i(M_1M_2)\ell_{III} \cdot \gamma^5\label{fac3}
\end{eqnarray}
Here, $c_W=(G_F/\sqrt{2})V_{bc}V_{ud}^\dagger$;
$\chi_+=a_1$ for transitions with a charged meson in the final state and
$\chi_-=a_2$ for transitions with a neutral meson in the final state;
$c_- = c_1 - c_2$, and $\ell^\pm_{I}$, $\ell^{P^+}_{II_a}$;
$\ell^{Q^+}_{II_a}$, $\ell^{D^+}_{II_b}$, $\ell^{Q^+}_{II_b}$, $\ell_{III}$
are  flavor coefficients whose values are listed in Table 1.
We employ the notation $M_\pm=M_1\pm M_2,$
$Q_+=(M_1+M_2)^2-M_3^2,$ \hspace*{0.5cm} $P_+=(M_2+M_3)^2-M_1^2,$
\hspace*{0.5cm} $D_+=(M_1+M_3)^2-M_2^2,$
$\omega=(M_1^2+M_2^2-M_3^2)/2M_1M_2$ where $M_1$, $M_2$ and $M_3$ are
the masses of the initial and final baryons, and the meson, respectively.

The contributions from diagrams IIb and III are down by
the helicity suppression  factor $(M_2/M_1)$  relative to the leading
diagrams I and IIa in agreement with the results of the spectator
model~\cite{Kramer}. Note that there are no any additional (dynamical) mass
suppression factors for the nonfactorizing diagrams in our approach.
In particular there is no mass suppression of diagram IIa. The overlap
factors $f(\omega)$, $H_2(\omega)$ and $H_3(\omega)$ pertain to the
contributions of diagrams I, II and III, respectively and are given
by
\begin{eqnarray}\label{overlap}
f(\omega)&=&\frac{R(\omega,\bar\Lambda)}{R(1,\bar\Lambda)}\\[2mm]
H_2(\omega)&=&t_2(r)\,
\frac{R_H(\omega,\bar\Lambda^{i},\bar\Lambda^{f})}
{\sqrt{R(1,\bar\Lambda^{i})R(1,\bar\Lambda^{f})}}
\,\frac{8}{9\pi\sqrt{3}}
\,\frac{\Lambda^4_{B_Q}}{\Lambda^3}\\[2mm]
H_3(\omega)&=&\frac{1}{2}\exp[9M_3^2/2\Lambda_{B_Q}^2]t_3(r)\,
\frac{R_H(\omega,\bar\Lambda^{i},\bar\Lambda^{f})}
{\sqrt{R(1,\bar\Lambda^{i})R(1,\bar\Lambda^{f})}}
\,\frac{8}{9\pi\sqrt{3}}
\,\frac{\Lambda^4_{B_Q}}{\Lambda^3}
\end{eqnarray}
where
\begin{eqnarray}
\hspace*{-.8cm}
R(\omega,\bar\Lambda)&=&\int\limits_0^\infty du u\int\limits_0^1d\alpha
\exp\biggl\{-18u^2 [1+2\alpha(1-\alpha)(\omega-1)]
+36u \bar\Lambda/\Lambda_{B_Q}\biggr\}
\nonumber\\
\hspace*{-.8cm}
R_{H}(\omega,\bar\Lambda^{i},\bar\Lambda^{f})
&=& \int\limits_0^\infty du u\int\limits_0^1d\alpha
\exp\biggl\{-72u^2[1+2\alpha(1-\alpha)(\omega-1)]\biggr\}
\nonumber\\
&\times&\exp\biggl\{144u(\bar\Lambda^{i}\alpha+\bar\Lambda^f(1-\alpha))/
\Lambda_{B_Q}-432u^2(\alpha^2+(1-\alpha)^2)\biggr\}
\nonumber
\end{eqnarray}
The parameters $\bar\Lambda^{i}$ and $\bar\Lambda^f$ are the
binding energy parameters of the initial and final baryons, respectively.
Terms proportional to $(M_1-M_2)/\Lambda_{B_Q}$ in the exponents have
been dropped for physical reasons. The quantities
$t_i(r)$ ($r=\Lambda/\Lambda_s$) are given by\\
\noindent
$t_2(r)=t_3(r)=(1+r)^2/4$ for
$\Lambda_b^0\to\Xi_c^0(\Xi_c^{\prime 0})+K^0$,
$\Xi_b^0\to\Lambda_c^+(\Sigma_c^+)+K^-$,\\
\noindent
$t_2(r)=(1+r)^2/4$ for
$\Xi_b^0\to\Omega_c^0+K^0$, $\Xi_b^-\to\Sigma_c^0+ K^-$,\\
\noindent
$t_3(r)=(1+r)^2/4$ for
$\Xi_b^0\to\Sigma_c^0+\bar K^0$,\\
\noindent
$t_2(r)=t_3(r)=r^2/(r^2\cos^2\delta_P+\sin^2\delta_P)^{1/2}$ for
$\Lambda_b^0\to\Sigma_c^0+\eta$,\\
\noindent
$t_2(r)=t_3(r)=r^2/(r^2\sin^2\delta_P+\cos^2\delta_P)^{1/2}$ for
$\Lambda_b^0\to\Sigma_c^0+\eta^\prime$,\\
\noindent
$t_i(r)=r^i/(r^2\cos^2\delta_P+\sin^2\delta_P)^{1/2}$ for
$\Xi_b^0\to\Xi_c^0(\Xi_c^{\prime 0})+\eta$,\\
\noindent
$t_i(r)=r^i/(r^2\sin^2\delta_P+\cos^2\delta_P)^{1/2}$ for
$\Xi_b^0\to\Xi_c^0(\Xi_c^{\prime 0})+\eta^\prime$,\\
\noindent
and $t_2(r)=t_3(r)=1$ for all other modes.
We use the notation $\delta_P=\theta_P-\theta_I$,
where $\theta_P=-11^{\rm o}$ is
the $\eta-\eta^\prime$ mixing angle, $\theta_I=35^{\rm o}$.

\section{Results}

In this section we give our numerical results for decay rates and
asymmetry parameters. The cutoff parameters $\Lambda_{B_q}$ and
$\Lambda_{B_Q}$, and
the binding energy parameter $\bar\Lambda$ are determined from
a fit to known branching ratios of nonleptonic decays
$\Lambda^+_c\to\Lambda^0\pi^+$, $\Lambda^+_c\to \Sigma^0\pi^+$,
$\Lambda^+_c\to \Sigma^+\pi^0$, $\Lambda^+_c\to p\bar K^0$
and $\Lambda^+_c\to \Xi^0 K^+$ (see, Table 2).
In the fit we use $\rho^2=1$ for the slope of the baryonic
Isgur-Wise function, leading to $\Lambda_{B_q}=3.037$ GeV,
$\Lambda_{B_Q}=2.408$ GeV, $\bar\Lambda$ = 0.9 GeV.
The parameters $\bar\Lambda_s$ and $\bar\Lambda_{ss}$
cannot be determined at present due to the lack of experimental
information on the decays of heavy-light baryons containing one
or two strange quarks. For the time being we fix them at the
values $\bar\Lambda_s$=1 GeV and $\bar\Lambda_{ss}$=1.1 GeV.
The masses of hadrons are taken from Ref.~\cite{PDG}.

In Table 3 we give our results for the decay rates and asymmetry
parameters for the exclusive nonleptonic $b\to c\bar u s$ decays
considered in this paper. A clear pattern emerges. The dominant
rates are into channels with factorizing contributions. Rates which
proceed only via nonfactorizing diagrams are small but not
negligibly small.

In Table 4 we give the contributions of nonfactorizing diagrams
relative to those of the factorizing ones for the decay
$\Lambda_b^0\to\Lambda_c^+\pi^-$ which we predict to have the
largest branching ratio. The values for overlap integrals for this
mode are $f=0.61$, $H_2$=24 MeV and $H_3$=12 MeV.
The suppressions of the contributions of diagrams IIb and III can in part
be traced to the helicity suppression factor.
The total contribution of the nonfactorizing diagrams can be seen to be
destructive. The sum of nonfactorizing contributions amount
up to 30~\% of the factorizing contribution in amplitude.
Using $\tau(\Lambda_b)=(1.14 \pm 0.08) \times 10^{-12}$ s \cite{PDG}
we predict a branching ratio of this mode of $(0.44\pm 0.003)\%$.
If one neglects the nonfactorizing contributions
for this mode as was done in~\cite{Cheng3} one would obtain an enhanced
rate of $\Gamma=0.665\times 10^{10} {\rm s}^{-1}$. The prediction for
the assymetry parameter remains at $\alpha\simeq -1$ and is thus not
affected by such an omission.

In conclusion, we have calculated the exclusive nonleptonic bottom to charm
baryon decays $\frac{1}{2}^+\to\frac{1}{2}^++0^-$ with a light pseudoscalar
meson in the final state. The dominant rates are into channels with
factorizing contributions. Decays which proceed only via the nonfactorizing
contributions occur at the 10~\% level of the modes with factorizing
contributions. In the decay $\Lambda_b^0\to\Lambda_c^+\pi^-$ the total
contribution of the nonfactorizing diagrams is destructive and amounts up
to 30~\% of the factorizing contributions in amplitude.
The generalization to the channels $\frac{1}{2}^+\to\frac{1}{2}^++1^-$,
$\frac{1}{2}^+\to\frac{3}{2}^++0^-$ and $\frac{1}{2}^+\to\frac{1}{2}^++1^-$
involving the ground state partners of the mesons and baryons in the final
state is straightforward and will be treated in a subsequent paper.
In this paper we discussed only the Cabibbo favoured decays induced by the
transitions $b\to c\bar u d$ with a light meson in the final state.
There are also a number of Cabibbo favoured decays with  heavy mesons
in the final  state which include the decays induced by the quark transitions
$ b\to c \bar c s$. The treatment of heavy mesons in the final state requires
some refinements in our simple Lagrangian spectator model. Again, exclusive
nonleptonic heavy baryon decays involving heavy mesons in the final state
are the subject of a future publication.

\vspace*{.3cm}
\noindent
{\bf Acknowledgments}
\vspace*{.5cm}

\noindent
M.A.I, V.E.L and A.G.R thank Mainz University for the hospitality
where a part of this work was completed.
This work was supported in part by the Heisenberg-Landau Program,
by the Russian Fund of Fundamental Research (RFFR) under contract
96-02-17435-a, the State Committee of the Russian Federation for
Education (project N 95-0-6.3-67, Grant Center at S.-Petersburg State
University) and  by the BMBF (Germany) under contract 06MZ566.
J.G.K. aknowledges partial support
by the BMBF (Germany) under contract 06MZ566.

\newpage
\noindent
{\bf List of Tables}
\vspace*{1cm}

\noindent
{\bf Table 1:} Flavour coefficients for exclusive
$b\rightarrow c\bar u d$  nonleptonic heavy baryon decays\\
($C\equiv \cos\delta_P$, $S\equiv \sin\delta_P$,
$\delta_P=\theta_P-\theta_I$, where $\theta_P=-11^{\rm o}$ is
the $\eta-\eta^\prime$ mixing angle, $\theta_I=35^{\rm o}$).

\noindent
{\bf Table 2:} Fit of the branchings of nonleptonic
decays of charm baryons (in \%).
Numerical values of CKM elements and Wilson coefficients:
$|V_{cs}|=0.975$, $|V_{ud}|=0.975$, $a_1=1.3$, $a_2=-0.65$.

\noindent
{\bf Table 3:} Decay rates and asymmetry parameters
for exclusive $b\rightarrow c\bar u d$ nonleptonic
heavy baryon decays. Numerical values of CKM elements and
Wilson coefficients:
$|V_{cb}|=0.04$, $|V_{ud}|=0.975$, $a_1=1.03$, $a_2=0.10$.

\noindent
{\bf Table 4:} Decay $\Lambda_b^0\to \Lambda_c^+\pi^-$:
Contributions of nonfactorizing diagrams  relative to
contributions of factorizing diagram.

\vspace*{4cm}
\noindent
{\bf List of Figures}
\vspace*{1cm}

\noindent
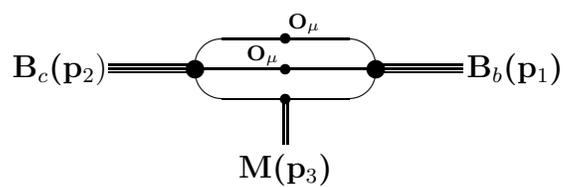
{\bf Figure 1:}
Nonleptonic decay of bottom baryon

\newpage
\hspace*{1.5cm}{\bf Table 1.}

\vspace*{.2cm}
\begin{center}
\def\arraystretch{1.5}
\begin{tabular}{|c|c|c|c|c|c|c|c|}
\hline\hline
Decay  & $\ell^-_{FD}$ & $\ell^+_{FD}$  & $\ell^{P_+}_{II_a}$
& $\ell^{Q_+}_{II_a}$  & $\ell^{D_+}_{II_b}$ & $\ell^{Q_+}_{II_b}$
& $\ell_{III}$  \\
\hline\hline
$\Lambda_b^0\to \Lambda_c^+\pi^-$& $-1$ & $-1$ & $-\frac{1}{2}$
& $\frac{1}{2}$ & $\frac{1}{2}$ & $\frac{1}{2}$ & $-2$ \\
\hline
$\Lambda_b^0\to \Sigma_c^+\pi^-$  & $0$ & $0$ & $\frac{\sqrt{3}}{2}$
& $\frac{1}{2\sqrt{3}}$ & $\frac{\sqrt{3}}{2}$ & $\frac{\sqrt{3}}{2}$
& $-2\sqrt{3}$ \\
\hline
$\Lambda_b^0\to \Sigma_c^0\pi^0$ & $0$ & $0$ & $-\frac{\sqrt{3}}{2}$
& $-\frac{1}{2\sqrt{3}}$ & $-\frac{\sqrt{3}}{2}$ & $-\frac{\sqrt{3}}{2}$
& $2\sqrt{3}$ \\
\hline
$\Lambda_b^0\to\Sigma_c^0\eta$ & $0$ & $0$
& $-\frac{\sqrt{3}}{2}S$ & $-\frac{1}{2\sqrt{3}}S$
& $\frac{\sqrt{3}}{2}S$ & $\frac{\sqrt{3}}{2}S$
& $2\sqrt{3}S$\\
\hline
$\Lambda_b^0\to\Sigma_c^0\eta^\prime$ & $0$ & $0$
& $\frac{\sqrt{3}}{2}C$ & $\frac{1}{2\sqrt{3}}C$
& $-\frac{\sqrt{3}}{2}C$ & $-\frac{\sqrt{3}}{2}C$
& $-2\sqrt{3}C$\\
\hline
$\Lambda_b^0\to \Xi_c^0 K^0$ & $0$ & $0$ & $\frac{1}{2}$ &
$-\frac{1}{2}$ & $0$ & $0$ & $2$ \\
\hline
$\Lambda_b^0\to \Xi_c^{\prime 0} K^0$ & $0$ & $0$ & $-\frac{\sqrt{3}}{2}$ &
$-\frac{1}{2\sqrt{3}}$ & $0$ & $0$ & $2\sqrt{3}$ \\
\hline
$\Xi_b^0\to \Xi_c^+\pi^-$  & $-1$ & $-1$ & $-\frac{1}{2}$
& $\frac{1}{2}$ & $0$ & $0$ & $0$ \\
\hline
$\Xi_b^0\to\Xi_c^{\prime +}\pi^-$ & $0$ & $0$
& $-\frac{\sqrt{3}}{2}$ & $-\frac{1}{2\sqrt{3}}$
& $0$ & $0$ & $0$\\
\hline
$\Xi_b^0\to \Xi_c^0\pi^0$ & $0$ & $0$ & $\frac{1}{2\sqrt{2}}$
& $-\frac{1}{2\sqrt{2}}$
& $\frac{1}{2\sqrt{2}}$ & $\frac{1}{2\sqrt{2}}$ & $0$ \\
\hline
$\Xi_b^0\to\Xi_c^0\eta$ & $0$ & $0$
& $\frac{1}{2\sqrt{2}}S$ & $-\frac{1}{2\sqrt{2}}S$
& $-\frac{1}{2\sqrt{2}}S$ & $-\frac{1}{2\sqrt{2}}S$
& $-2C$\\
\hline
$\Xi_b^0\to\Xi_c^0\eta^\prime$ & $0$ & $0$
& $-\frac{1}{2\sqrt{2}}C$ & $\frac{1}{2\sqrt{2}}C$
& $\frac{1}{2\sqrt{2}}C$ & $\frac{1}{2\sqrt{2}}C$
& $-2S$\\
\hline
$\Xi_b^0\to\Xi_c^{\prime 0}\pi^0$ & $0$ & $0$ & $\frac{\sqrt{3}}{2}$ &
$\frac{1}{2\sqrt{3}}$ & $\frac{\sqrt{3}}{2}$ & $\frac{\sqrt{3}}{2}$ & $0$ \\
\hline
$\Xi_b^0\to\Xi_c^{\prime 0}\eta$ & $0$ & $0$
& $\frac{\sqrt{3}}{2\sqrt{2}}S$
& $\frac{1}{2\sqrt{6}}S$
& $-\frac{\sqrt{3}}{2\sqrt{2}}S$
& $-\frac{\sqrt{3}}{2\sqrt{2}}S$
& $-2\sqrt{3}C$\\
\hline
$\Xi_b^0\to\Xi_c^{\prime 0}\eta^\prime$ & $0$ & $0$
& $-\frac{\sqrt{3}}{2\sqrt{2}}C$
& $-\frac{1}{2\sqrt{6}}C$
& $\frac{\sqrt{3}}{2\sqrt{2}}C$
& $\frac{\sqrt{3}}{2\sqrt{2}}C$
& $-2\sqrt{3}S$\\
\hline
$\Xi_b^0\to\Lambda_c^+ K^-$ & $0$ & $0$ & $0$ & $0$
& $-\frac{1}{2}$ & $-\frac{1}{2}$ & $2$ \\
\hline
$\Xi_b^0\to\Sigma_c^+ K^-$ & $0$ & $0$ & $0$ & $0$
& $-\frac{\sqrt{3}}{2}$ & $-\frac{\sqrt{3}}{2}$
& $2\sqrt{3}$\\
\hline
$\Xi_b^0\to\Sigma_c^0\bar K^0$ &$0$ &$0$ &$0$ &$0$ & $0$ & $0$&$-2\sqrt{6}$\\
\hline
$\Xi_b^0\to\Omega_c^0 K^0$ & $0$ & $0$ & $\sqrt{\frac{3}{2}}$
& $\frac{1}{\sqrt{6}}$ & $0$ & $0$ & $0$\\
\hline
$\Xi_b^-\to \Xi_c^0 \pi^-$ & $-1$ & $-1$
& $\frac{1}{2}$ & $\frac{1}{2}$ & $0$ & $0$ & $0$\\
\hline
$\Xi_b^-\to \Xi_c^{\prime 0}\pi^-$ & $0$ & $0$
& $0$ & $0$ & $\frac{\sqrt{3}}{2}$ & $\frac{\sqrt{3}}{2}$ & $0$ \\
\hline
$\Xi_b^-\to \Sigma_c^0 K^-$ & $0$ & $0$
& $0$ & $0$ & $-\sqrt{\frac{3}{2}}$ & $-\sqrt{\frac{3}{2}}$ & $0$ \\
\hline
$\Omega_b^-\to \Omega_c^0\pi^-$& $-1$ & $ \frac{1}{3}$
& $0$ & $0$& $0$ & $0$ & $0$\\
\hline\hline
\end{tabular}
\end{center}

\newpage
\hspace*{4cm}{\bf Table 2.}

\vspace*{.2cm}
\def\arraystretch{1.5}
\begin{center}
\begin{tabular}{|c|c|c|}
\hline\hline
Process & \,\,\,\,\, Our fit \,\,\,\,\, & \,\,\,Experiment~\cite{PDG}\,\,\, \\
\hline
$\Lambda^+_c\to \Lambda \pi^+$ & 0.79 & 0.79$\pm$ 0.18\\
\hline
$\Lambda^+_c\to \Sigma^0 \pi^+$ & 0.88 & 0.88$\pm$ 0.20\\
\hline
$\Lambda^+_c\to \Sigma^+ \pi^0$ & 0.88 & 0.88$\pm$ 0.22\\
\hline
$\Lambda^+_c\to p \bar K^0$ & 2.06 & 2.2$\pm$ 0.4\\
\hline
$\Lambda^+_c\to \Xi^0 K^+$ & 0.31 & 0.34$\pm$ 0.09\\
\hline\hline
\end{tabular}
\end{center}

\vspace*{1cm}

\noindent
\hspace*{1.5cm}{\bf Table 3.}

\vspace*{.2cm}
\begin{center}
\def\arraystretch{1.5}
\begin{tabular}{|c|c|c||c|c|c|}
\hline\hline
Process  & $\Gamma$ (in 10$^{10}$ s$^{-1}$) & $\alpha$  &
Process  & $\Gamma$ (in 10$^{10}$ s$^{-1}$) & $\alpha$  \\
\hline\hline
$\Lambda_b^0\to \Lambda_c^+\pi^-$     & 0.382 & -0.99 &
$\Xi_b^0\to\Xi_c^{\prime 0}\pi^0$ & 0.014 & 0.94\\
\hline
$\Lambda_b^0\to \Sigma_c^+\pi^-$      & 0.039 & 0.65 &
$\Xi_b^0\to\Xi_c^{\prime 0}\eta$ & 0.015 & -0.98\\
\hline
$\Lambda_b^0\to \Sigma_c^0\pi^0$      & 0.039 & 0.65 &
$\Xi_b^0\to\Xi_c^{\prime 0}\eta^\prime$& 0.021 & 0.97\\
\hline
$\Lambda_b^0\to\Sigma_c^0\eta$        & 0.023 & 0.79 &
$\Xi_b^0\to\Lambda_c^+ K^-$ & 0.010 & -0.73\\
\hline
$\Lambda_b^0\to\Sigma_c^0\eta^\prime$ & 0.029 & 0.99&
$\Xi_b^0\to\Sigma_c^+ K^-$ & 0.030 & -0.74\\
\hline
$\Lambda_b^0\to \Xi_c^0 K^0$          & 0.021 & -0.81 &
$\Xi_b^0\to\Sigma_c^0\bar K^0$ & 0.021 & 0\\
\hline
$\Lambda_b^0\to \Xi_c^{\prime 0} K^0$ & 0.032 & 0.98 &
$\Xi_b^0\to\Omega_c^0 K^0$ & 0.023 & 0.65\\
\hline
$\Xi_b^0\to \Xi_c^+\pi^-$ & 0.479 & -1.00 &
$\Xi_b^-\to \Xi_c^0 \pi^-$ & 0.645 & -0.97\\
\hline
$\Xi_b^0\to\Xi_c^{\prime +}\pi^-$& 0.018 & 0.61 &
$\Xi_b^-\to \Xi_c^{\prime 0}\pi^-$& 0.007 &-1.00\\
\hline
$\Xi_b^0\to \Xi_c^0\pi^0$ & 0.002 & -0.99 &
$\Xi_b^-\to \Sigma_c^0 K^-$ & 0.016 & -0.98\\
\hline
$\Xi_b^0\to\Xi_c^0\eta$ & 0.012 & -0.86 &
$\Omega_b^-\to \Omega_c^0\pi^-$ & 0.352 & 0.60\\
\hline
$\Xi_b^0\to\Xi_c^0\eta^\prime$ & 0.003 & 0.71 & & &\\
\hline\hline
\end{tabular}
\end{center}

\newpage

\noindent
\hspace*{4cm}{\bf Table 4.}

\vspace{0.5cm}

\begin{center}
\def\arraystretch{2.}
\begin{tabular}{|c|c|c|c|c|}
\hline\hline
Amplitude   & \multicolumn{4}{|c|} {Diagram} \\
\cline{2-5}  & ${\rm IIa}$  & ${\rm IIb}$ & ${\rm IIa}+{\rm IIb}$ &
${\rm III}$ \\
\hline\hline
A   & -13.9$\%$ & -6.2$\%$ & -20.1$\%$ & \\
\hline
B   & -14.3$\%$ & -5.8$\%$ & -20.1$\%$ &  -8.5$\%$  \\
\hline\hline
\end{tabular}
\end{center}

\newpage

\unitlength=0.50mm
\special{em:linewidth 0.4pt}
\linethickness{0.4pt}
\hspace*{4cm}
\begin{picture}(127.00,116.00)
\put(70.00,23.00){\oval(48.00,16.00)[]}
\put(45.00,23.00){\circle*{5.00}}
\put(95.00,23.00){\circle*{5.00}}
\put(95.00,23.00){\line(1,0){25.00}}
\put(120.00,24.00){\line(-1,0){25.00}}
\put(95.00,22.00){\line(1,0){25.00}}
\put(45.00,22.00){\line(-1,0){25.00}}
\put(20.00,23.00){\line(1,0){25.00}}
\put(45.00,24.00){\line(-1,0){25.00}}
\put(44.00,24.00){\line(3,5){26.00}}
\put(96.00,24.00){\line(-3,5){26.00}}
\put(70.00,67.00){\circle*{3.00}}
\put(6.00,23.00){\makebox(0,0)[cc]{{\bf B$_c$(p$_2$)}}}
\put(134.00,23.00){\makebox(0,0)[cc]{{\bf B$_b$(p$_1$)}}}
\put(70.00,78.00){\circle{14.00}}
\put(70.00,85.00){\circle*{3.00}}
\put(70.00,71.00){\circle*{3.00}}
\put(70.50,85.00){\line(0,1){15.00}}
\put(69.50,85.00){\line(0,1){15.00}}
\put(70.00,108.00){\makebox(0,0)[cc]{{\bf M(p$_3$)}}}
\put(57.00,74.00){\makebox(0,0)[cc]{{\scriptsize\bf O$_\mu$}}}
\put(80.00,66.00){\makebox(0,0)[cc]{{\scriptsize\bf O$_\mu$}}}
\end{picture}

\begin{center}
\hspace*{1cm}I
\end{center}

\vspace*{-2.25cm}
\begin{picture}(127.00,116.00)
\put(70.00,23.00){\oval(48.00,16.00)[]}
\put(46.00,23.00){\circle*{5.00}}
\put(94.00,23.00){\circle*{5.00}}
\put(95.00,23.00){\line(1,0){20.00}}
\put(95.00,24.00){\line(1,0){20.00}}
\put(95.00,22.00){\line(1,0){20.00}}
\put(45.00,23.00){\line(1,0){50.00}}
\put(45.00,22.00){\line(-1,0){20.00}}
\put(45.00,23.00){\line(-1,0){20.00}}
\put(45.00,24.00){\line(-1,0){20.00}}
\put(11.00,23.00){\makebox(0,0)[cc]{{\bf B$_c$(p$_2$)}}}
\put(130.00,23.00){\makebox(0,0)[cc]{{\bf B$_b$(p$_1$)}}}
\put(70.00,31.00){\circle*{3.00}}
\put(80.00,15.00){\circle*{3.00}}
\put(62.00,15.00){\circle*{3.00}}
\put(61.50,15.00){\line(0,-1){12.00}}
\put(62.50,15.00){\line(0,-1){12.00}}
\put(70.00,37.00){\makebox(0,0)[cc]{{\scriptsize\bf O$_\mu$}}}
\put(78.00,9.00){\makebox(0,0)[cc]{{\scriptsize\bf O$_\mu$}}}
\put(62.00,-3.80){\makebox(0,0)[cc]{{\bf M(p$_3$)}}}
\end{picture}
\hspace*{1cm}
\begin{picture}(127.00,116.00)
\put(70.00,23.00){\oval(48.00,16.00)[]}
\put(46.00,23.00){\circle*{5.00}}
\put(94.00,23.00){\circle*{5.00}}
\put(95.00,23.00){\line(1,0){20.00}}
\put(95.00,24.00){\line(1,0){20.00}}
\put(95.00,22.00){\line(1,0){20.00}}
\put(45.00,23.00){\line(1,0){50.00}}
\put(45.00,22.00){\line(-1,0){20.00}}
\put(45.00,23.00){\line(-1,0){20.00}}
\put(45.00,24.00){\line(-1,0){20.00}}
\put(11.00,23.00){\makebox(0,0)[cc]{{\bf B$_c$(p$_2$)}}}
\put(130.00,23.00){\makebox(0,0)[cc]{{\bf B$_b$(p$_1$)}}}
\put(70.00,31.00){\circle*{3.00}}
\put(78.00,15.00){\circle*{3.00}}
\put(62.00,15.00){\circle*{3.00}}
\put(77.50,15.00){\line(0,-1){12.00}}
\put(78.50,15.00){\line(0,-1){12.00}}
\put(70.00,37.00){\makebox(0,0)[cc]{{\scriptsize\bf O$_\mu$}}}
\put(61.00,9.00){\makebox(0,0)[cc]{{\scriptsize\bf O$_\mu$}}}
\put(78.00,-3.80){\makebox(0,0)[cc]{{\bf M(p$_3$)}}}
\end{picture}

\begin{center}
\vspace*{1cm}
\hspace*{-1cm} IIa  \hspace*{6cm} IIb
\end{center}

\vspace*{-2.25cm}
\hspace*{3cm}
\begin{picture}(127.00,116.00)
\put(70.00,23.00){\oval(48.00,16.00)[]}
\put(46.00,23.00){\circle*{5.00}}
\put(94.00,23.00){\circle*{5.00}}
\put(95.00,23.00){\line(1,0){22.00}}
\put(95.00,24.00){\line(1,0){22.00}}
\put(95.00,22.00){\line(1,0){22.00}}
\put(45.00,23.00){\line(1,0){50.00}}
\put(45.00,22.00){\line(-1,0){22.00}}
\put(45.00,23.00){\line(-1,0){22.00}}
\put(45.00,24.00){\line(-1,0){22.00}}
\put(10.00,23.00){\makebox(0,0)[cc]{{\bf B$_c$(p$_2$})}}
\put(131.00,23.00){\makebox(0,0)[cc]{{\bf B$_b$(p$_1$)}}}
\put(70.00,31.00){\circle*{3.00}}
\put(70.00,23.00){\circle*{3.00}}
\put(70.00,15.00){\circle*{3.00}}
\put(70.50,15.00){\line(0,-1){12.00}}
\put(69.50,15.00){\line(0,-1){12.00}}
\put(75.00,35.00){\makebox(0,0)[cc]{{\scriptsize\bf O$_\mu$}}}
\put(64.00,26.50){\makebox(0,0)[cc]{{\scriptsize\bf O$_\mu$}}}
\put(70.00,-3.80){\makebox(0,0)[cc]{{\bf M(p$_3$)}}}
\end{picture}
\vspace*{1cm}

\begin{center}
III
\end{center}

\begin{center}
Fig.1
\end{center}
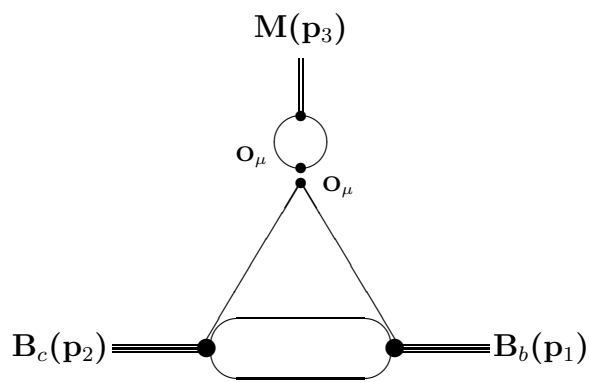
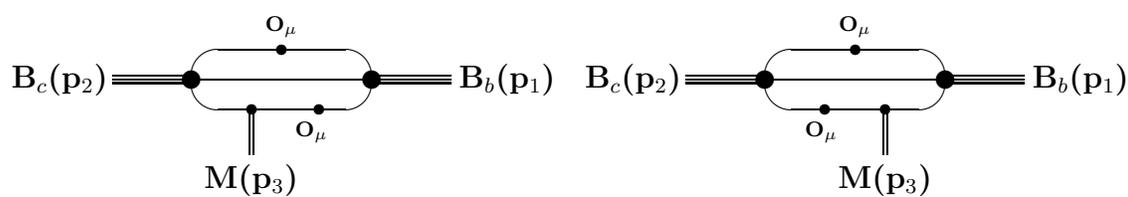
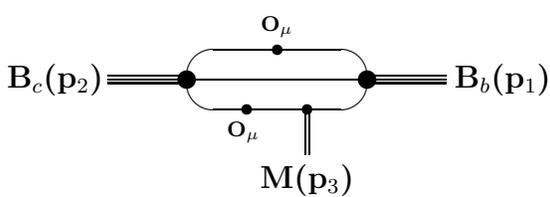
\end{document}